\documentclass[sigconf,nonacm,hyphens]{acmart}
\usepackage{balance} 
\usepackage{cite}

\usepackage{amsmath,amssymb,amsfonts}
\usepackage{algorithmic}
\usepackage{graphicx}
\usepackage{textcomp}
\usepackage{xcolor}
\usepackage{caption}
\usepackage{subcaption}
\usepackage{float}
\usepackage{url}
\usepackage{natbib}
\begin{document}

\title{The CROSS Incubator: \break A Case Study for funding and training RSEs}

\author{Stephanie \break Lieggi}
\affiliation{%
  \institution{UC Santa Cruz}
}
\email{slieggi@ucsc.edu}

\author{Ivo \break Jimenez}
\affiliation{%
  \institution{UC Santa Cruz}
}
\email{ivotron@ucsc.edu}

\author{Jeff \break LeFevre}
\affiliation{%
  \institution{UC Santa Cruz}
}
\email{jlefevre@ucsc.edu}

\author{Carlos \break Maltzahn}
\affiliation{%
  \institution{UC Santa Cruz}
}
\email{carlosm@ucsc.edu}

\renewcommand{\shortauthors}{Lieggi, et al.}

\maketitle


\section{Introduction} \label{sec:introduction}
The incubator and research projects sponsored by the Center for Research in Open Source Software (CROSS, \url{cross.ucsc.edu}) at UC Santa Cruz have been very effective at promoting the professional and technical development of research software engineers. Carlos Maltzahn founded CROSS in 2015 with a generous gift of \$2,000,000 from UC Santa Cruz alumnus Dr.~Sage Weil~\citep{ucsc-news} and founding memberships of Toshiba America Electronic Components, SK Hynix Memory Solutions, and Micron Technology. Over the past five years, CROSS funding has enabled PhD students to not only create  research software projects but also learn how to draw in new contributors and leverage established open source software communities. 
This position paper will present CROSS fellowships as case studies for how university-led open source projects can create a real-world, reproducible model for effectively training, funding and supporting research software engineers.
\section{CROSS}\label{sec:two}
The Center for Research in Open Source Software bridges the gap between research software created by students and successful open source software projects. As part of the UC Santa Cruz Baskin School of Engineering, it is sustained by industry memberships and provides fellowships across multiple research groups in multiple departments.

\subsection{Research and Incubator Fellows}
CROSS supports the work of two types of fellows: a research fellow is a PhD student who is selected for their research and whose research software presents a plausible path to a successful open source software project. An incubator fellow is a post-doctoral appointment with the goal to grow a contributor community around open source research software. Once selected, all fellows are reviewed twice a year. Positive reviews extend funding as long as funds are available and until the program's goal is achieved: for research fellows that is graduation; for incubator fellows that is some form of independent sustainability (typically within a four year limit.) 


\subsection{The Open Source Research Experience}

CROSS found that research and incubator projects benefit the educational and professional development of a wide range of students, lowering the barriers to participation and helping them to productively engage in open source communities. Motivated by their programmatic requirements, including class, senior, capstone, summer, or master projects, combined with the excellent mentorship of CROSS fellows, graduate and undergraduate students have actively contributed to CROSS projects at various levels and became important members of the contributor communities around these projects~\citep{oram}.

 
 The positive experience with student contributors, both local and global via CROSS' multi-year participation in the Google Summer of Code~\citep{gsoc} motivated the formation of the CROSS Open Source Research Experience (OSRE) which acts as a marketplace for open source project ideas~\citep{osre}. This marketplace matches students with mentors and allows industry to engage by funding successful matches (a form of industry engagement that is expected to increase in importance during the pandemic, for example~\citep{google-intern}).  
\section{RSE skills taught by CROSS}\label{sec:sec3}
CROSS embeds fellows in a concentrated environment of mentorship by top experts in research, industry, and open source foundations. The CROSS advisory committee consists of world-class experts in open source software and entrepreneurship, including Doug Cutting (Chief Architect at Cloudera), Karen Sandler (Executive Director at Software Freedom Conservancy), Nissa Strottman (VP Products \& Operations at VISA), Sage Weil (Ceph Principal Architect at Red Hat), and James Davis (Professor at UC Santa Cruz). The committee, together with the members of the Industrial Advisory Board who have years of experience at translating university research into product research and development, provide mentorship in the following three areas: open source software project leadership, mentoring and training of code contributors and student researchers, and industry engagement toward project uptake and funding opportunities.


\paragraph{Open Source Software Project Leadership} Incubator fellows (and to some extent research fellows) lead their projects in an open source environment and interact with the local and broader student community as well as industry partners.
This motivates them to develop skills in open source project management and community engagement, through learning and applying best practices in the dynamic open source world.
Incubator fellows navigate the task of taking a project from research prototype status toward more robust and viable production quality releases while learning how to leverage existing open source software projects and their communities to avoid any software maintenance bottlenecks.


\paragraph{Mentoring and Training} A key element of professional development for RSEs is outreach and training of students and other contributors, both of which may come from varying capability levels in their careers.
Students and contributors are recruited through campus communications, classes, or other programs that fellows participate in, e.g. GSoC or the NSF-funded Institute for Research and Innovation in Software for High-Energy Physics (IRIS-HEP)~\citep{iris-hep}.
CROSS fellows advertise project ideas via the OSRE marketplace, review applications, interact with the applicants to help develop their proposals into viable projects for selection and completion within the given time frame (typically 3 months). 

Projects with user-friendly documentation as well as ease of onboarding for new contributors are much more likely to succeed in the OSRE marketplace and in the real world. One example of this is through the use of Popper (a CROSS incubator project itself; see §\,\ref{sec:popper}), a container native execution and workflow automation engine, which for our incubator case can simplify the task of compiling and testing software projects. This is helpful for new contributors at all levels to quickly get them up and running, and thus able to begin making contributions to the project itself rather than spending time on things such as build environments and dependency issues.

Both incubator and research fellows are supervising multiple local, national, and international students at the B.S., M.S., and Ph.D. levels, gaining valuable people and priority management experience which are among the most challenging aspects not only of mentoring and training but also of project leadership.
Understanding individual student goals and interests in order to keep them engaged in the project is part of the mentoring process that is both critical and rewarding in an open source environment that is largely based on volunteer activity. 


\paragraph{Fundraising} 
Worldwide, CROSS is among very few university open source software efforts which are able to raise significant funding from industry, government, and private donations based on their open source software project portfolio. Incubator fellows in collaboration with CROSS leadership and partners in industry, academia, and government are constantly engaged in fundraising. They learn how to successfully apply for outside fellowships to fund student contributors and to increase funding of their projects via governmental grants and consulting opportunities. The CROSS incubator program, with its network of advisors and partners, provides great fundraising mentorship and valuable pointers to outside funding opportunities. Turning this support into successful open source projects in turn attracts both talent and long-term financing.

\section{Incubators Projects: Case Studies}

Incubator fellows grow communities around their projects by onboarding and mentoring contributors and maintaining project web pages, documentation, tutorials, source code repositories, forums on various platforms such as email lists and Slack, regular developer meetings, and up-to-date project ideas on the OSRE web page for which students can apply.
Fellows also participate in research meetings within the relevant labs at the university, which becomes a good opportunity for mentoring graduate-level students and keeping the projects aligned with the latest research developments.  
Research groups often include students at external universities and researchers at several national labs that collaborate on projects and research papers and often end up becoming important contributors.

Fellows interact with CROSS industry partners and advisory board members during regular project review sessions and through our annual research symposium, where students present posters or talks on their contributions, and fellows have the opportunity to organize and offer tutorials and interactive topical sessions with attendees from industry, government, and academia. 

CROSS incubator projects provide useful case studies for the type of programs that can foster the development of skills relevant for RSEs. Two of our current incubator projects --  Black Swan/Popper and Skyhook Data Management -- are highlighted below.

\paragraph{Popper}
\label{sec:popper}
The "Black Swan: The Popper Reproducibility Platform" incubator (\url{https://getpopper.io/}) aims to create infrastructure for reproducible open science. This incubator has helped the development of skills related to RSEs, such as when the incubator fellow has led the organization and instruction of hands-on tutorials targeted to scientists from multiple domains. Through these experiences, the fellow has been able to interact with users and document their experiences, thus improving the usability aspects of the software that is being developed as part of this project. In general, incubator projects present the opportunity to learn about community-management aspects of a project.
Another related, formative aspect of the program is that all CROSS incubator projects make use of existing OSS, which in practice means interacting with the communities associated to these upstream projects. This presents an opportunity for fellows and undergraduate students to gain experience working with wider more diverse OSS communities.

\paragraph{SkyhookDM}
The Skyhook Data Management project (\url{https://www.skyhookdm.com}) offers storage and management of tabular data within Ceph~\citep{weil2006} distributed object storage.  
SkyhookDM leverages IO and CPU power on storage servers by developing library extensions to Ceph that perform data management tasks for query processing and physical design at the object level. 
The incubator fellow on this project had existing research experience, but has gained soft-skills experience in managing projects and people, including supervising graduate class projects, Master's projects, and supervising undergraduate programming assistants as part of the CROSS OSRE program.
In particular, undergraduate students have gained valuable experience in software development for large, complex projects, which is quite different from classroom project programming experience, as has been noted by several student programmers.

A specific challenge has been learning and adapting to deployment techniques for cloud-scale, which is a requirement for wider adoption of this project.
The SkyhookDM project straddles the boundary between databases and storage systems, providing new opportunities for data management, and it has given the incubator fellow the opportunity to be part of industry discussions regarding the move toward computational storage devices.

\section{Conclusion}\label{sec:conclusion}

CROSS activities, particularly incubator projects like SkyhookDM and Popper, provide clear examples of how to develop RSEs' skills in a university environment, through collaboration with experts in industry, government and academia. The CROSS incubator model is financially sustainable, with industry and other outside sponsors supporting the work of CROSS incubator fellows. CROSS provides a  model for how students and postdocs, through effective mentorship, build a contributor community for their projects and train the next generation of RSEs. 

\bibliography{bibfiles/bibliography.bib}

\end{document}